\documentclass[aps,prd,twocolumn]{revtex4-1}
\usepackage[english]{babel} 
\usepackage{amssymb}
\usepackage{amsmath}
\usepackage{txfonts}
\usepackage{mathdots}
\usepackage{graphicx}
\usepackage{subfigure}
\usepackage{bm}

\newcommand{\hbrn}{\hat{\rm{\bf n}}}
\newcommand{\hbru}{\hat{\rm{\bf u}}}
\newcommand{\hbrv}{\hat{\rm{\bf v}}}
\newcommand{\hbrx}{\hat{\rm{\bf x}}}
\newcommand{\hbry}{\hat{\rm{\bf y}}}
\newcommand{\hbrz}{\hat{\rm{\bf z}}}
\newcommand{\Y}{\Y_{lm}}

\newcommand{\be}{\begin{equation}}
\newcommand{\ee}{\end{equation}}
\newcommand{\Be}{\begin{eqnarray}}
\newcommand{\Ee}{\end{eqnarray}}

\newcommand{\f}{\frac}

\begin{document}

\pagestyle{plain}

\title{Gravitational Waves by Perturbation of a Slowly Rotating Thin-Shell Wormhole
}

\author{Sung-Won Kim}
\email[email:]{sungwon@ewha.ac.kr}
\affiliation{Ewha Womans University, Seoul 03760,
Korea}

\begin{abstract}
In this paper, the gravitational wave generation by a slowly rotating thin-shell wormhole is considered. Since the rotating thin-shell wormhole is assumed to be an axisymmetric rigid body, the rotation axis coincides with the largest principal axis which means there is no precession motion.
However, if there is a perturbation in the angular velocity,
the rotating wormhole can move with precession by perturbation which make arise the gravitational waves. We derive the gravitational wave spectrum, energy loss rate, and angular momentum loss rate.
\end{abstract}

\maketitle
\date{today}
\section{Introduction}

The detections of the gravitational waves by the LIGO/Virgo collaboration \cite{Aasi,Ace} are in great success of windows to researches of new area. The improving accuracy and extending the detection bands
provide the motivations for exploring new gravitational sources. Wormhole also can be an astrophysical compact object which is one of the candidates of gravitational wave source,
even though their existence is still not confirmed yet.
However, considering the gravitational waves by wormholes has sufficient meaning
to be one of the candidates of source for future detections \cite{TBBL2021}.

There are several trials to find any footprint of wormhole, such as gravitational lensing \cite{ATKAA}, shadows \cite{NTY2013},  Einstein rings \cite{THY2012}, and particle creation \cite{SWK92}. If we succeed in detection of GW generated by wormhole, it also can be added to the list of wormhole evidence.

To get gravitational waves by a single stellar object, we need the periodic motion of the perturbed term in it. Periodic motion can mainly be caused by two dynamic motions: pulsation and precession.
For the case of pulsation, the perturbation is required in matter as well as in geometric part. The another dynamic motion is precession derived from the rotation of the body. In latter case, we use the inertia tensor of the rigid body instead of quadrupole moment. These two quantities (inertia tensor and quadrupole moment) are identified except the change of the sign, after the second derivation with respect to time in deriving the gravitational wave forms.

Usually two mathematical models of wormhole are used because of their simplicity: Ellis-Bronnikov-Morris-Thorne wormhole and Visser thin-shell wormhole.
Ellis \cite{Ellis} suggested a very simple drainhole model in early times. In his work, Bronnikov \cite{Bron} realized, with evidence, that the Ellis drainhole is geodesically complete, without event horizons, with free singularity and with traversability. Morris and Thorne \cite{MT} suggested a simple Schwarzschild type traversable wormhole, which is mathematically identical to the transformed Ellis one.
Visser \cite{Visser} tried to construct wormhole with minimized use of exotic matter by cut-and-paste operation with two copies of Schwarzschild spacetime. He used only delta function distribution of the exotic matter at the junction of two spacetimes, so that the resultant wormhole is the thin-shell wormhole.

As the first step to the study of the gravitational wave generated by wormhole, we considered the rotating thin-shell wormhole because of the minimal use of exotic matter.
One of the most challenging problem in Einstein's general relativity is to find stable wormholes with a minimum amount of exotic matter or with completely normal matter \cite{MT, HV}. In this regard we note that rotating wormholes present more alternatives because of their extra degrees of freedom \cite{Ov}. Rotating wormholes were first introduced by Teo \cite{Teo}, who
addressed the restrictions that must be imposed on the geometry in order to have a stationary and axisymmetric wormhole. However, the model did not think the matter part of the r.h.s of Einstein equation.
There is other type of rotating wormhole supported by a phantom field, slowly rotating case \cite{KS2008}, rapidly rotating case \cite{KK2014}, nonsymmetric case \cite{CKK2016}, and thin-shell wormhole \cite{Visser_book,KS11,Ov}.
The rotating thin-shell wormhole is made by two copies of Kerr black holes \cite{Visser_book,KS11} with angular momentum and inhomogeneous surface density following the mechanism designed by Visser.



In this paper we consider the gravitational wave generation by perturbed precession of thin-shell rotating wormhole.
Free precession happens when triaxial ellipsoid rotates or the principal axis of an axisymmetric body does not coincide with the angular momentum \cite{Mechanics}. If the axisymmetric body rotates around the principal axis, the body does not move with free precession. However, the perturbation in angular velocity of the principal axis causes oscillation and precession of axis like the forced motion. In the astrophysical situations, there is a high probability of perturbation of the star due to the scattering of small bodies.

\section{Thin-shell wormhole}

As the example of the gravitational wave generation, we consider the thin-shell model
according to Visser. Thin-shell wormhole is the surgical cut-and-paste operation with two identical copies of Schwarzschild black hole. The junction is located at the outer of the event horizon so that two way travel is possible.
To construct the rotating thin-shell wormhole, we combine two copies of Kerr black holes \cite{KS11}, at some place outside of event horizon like the non-rotating case \cite{Visser}.
Einstein equations on the shell is (Lanczos equations)
\be
-k_{ij}+kg_{ij}=8\pi S_{ij}
\ee
with the extrinsic curvature
\Be
k_{ij}&=& K^+_{ij}-K^-_{ij},~~k=k^i_i, \nonumber \\
K^{\pm }_{ij}&=& -n^\pm_\gamma \left.\left(\f{{\partial }^2x^{\gamma }}{\partial {\xi }^i\partial {\xi }^j}+\Gamma^{\gamma}_{\alpha\beta}\f{\partial x^{\alpha }}{\partial {\xi }^i}\frac{\partial x^{\beta }}{\partial {\xi }^j}\right)\right|_\Sigma,
\Ee
and the surface stress-energy tensor
\be
S_{ij}=\left( \begin{array}{ccc}
\sigma  & 0 & \zeta  \\
0 & p_{\vartheta } & 0 \\
\zeta  & 0 & p_{\varphi } \end{array}
\right),
\ee
where $\sigma $ is the surface energy density, $p_{\vartheta }$ and $p_{\varphi }$ are principal surface pressures, and $\zeta$ is the surface angular momentum density.

Kerr in Boyer-Lindquist coordinates is given by
\begin{widetext}
\be
ds^2 = -\left(1-\frac{2{\cal M}r}{{\rho}^2}\right)c^2dt^2+
\f{\rho^2}{\Delta}dr^2+\rho^2 d\theta^2
+ \left( r^2 + \rho^2 + \f{2{\cal M}a^2r}{\rho^2}\sin^2\theta \right) \sin^2\theta d\phi^2 - \f{4{\cal J}r}{\rho^2} \sin^2\theta d\phi c dt, \\
\label{BL}
\ee
\end{widetext}
where
\[
\rho^2=r^2+a^2\sin^2\theta, \quad \Delta =r^2-2{\cal M}r+a^2.
\]
Here ${\cal J}$ is the angular momentum in length-square dimension, and ${\cal M}$ is the mass in  length dimension. They have
\[
{\cal M}=M\times \left(\f{G}{c^2}\right), \qquad {\cal J}=J \times \left(\f{G}{c^3}\right),
\]
where $J$ is the angular momentum, $a={\cal J}/{\cal M}$ is the angular momentum per unit mass and it has the unit of length. The event horizon is located at $r_+ = {\cal M}+\sqrt{{\cal M}^2-a^2}$ and the ergosphere is between $r_+$ and $r_0 = {\cal M}+\sqrt{{\cal M}^2-a^2\cos^2\theta}$.

In this thin shell wormhole, the mass surface density is derived as \cite{KS11}
\be
4\pi \sigma =-\frac{{{\Delta}}^{1/2}_{\beta}[2{\beta}^3+{\alpha}^2\beta +{\alpha}^2+{\alpha}^2\left(\beta -1\right)\cos^2\theta ]}{{\cal M}{\rho}_{\beta}\Phi}\left(\frac{c^2}{G}\right),
\ee
where $\alpha =a/{\cal M}$, $\beta =b /{\cal M}$, $\Delta_\beta = \beta^2-2\beta +\alpha^2$,
$\rho^2_\beta =\beta^2 +\alpha^2 \cos^2\theta$, $\Phi=\beta^4+\alpha^2\beta^2+2\alpha^2\beta +\alpha^2\Delta_\beta \cos^2\theta $  which are all dimensionless parameter.
The $a$ is angular momentum per mass in length unit and $b$ is
the radius of cut-and-paste, that is, the wormhole throat size.
Thus for the thin-shell wormhole model, the event horizon is at ${\beta }_+=1+\sqrt{1-{\alpha }^2}$ and $1< \beta_+ <2 $ for $\alpha^2<1$.
At event horizon, $\sigma =0$.
The ergosphere is at ${\beta }_0=1+\sqrt{1-{\alpha }^2{{\mathrm{cos}}^2 \theta \ }}>{\beta }_+>1$. At least, $\beta > \beta_+$ for positivity of $\Delta_\beta$.
For the case of vanishing $\alpha$, then the density is
\be
\sigma_0 =-\frac{1}{2\pi b}\sqrt{1-2/\beta}\left(\frac{c^2}{G}\right).
\label{thin-density}
\ee
The density is negative everywhere, there is no region of positive value, even when the wormhole is rotating. Fig. 1 shows the $\theta$-distribution of density at constant $\alpha$ and $\beta$.
The property of $\sigma < 0$ is well discussed in the article by Morris and Thorne \cite{MT} in the context of flare-out condition.


\begin{figure}[t]
  \centering
    \subfigure
    {%
    \includegraphics[width=0.4\textwidth]{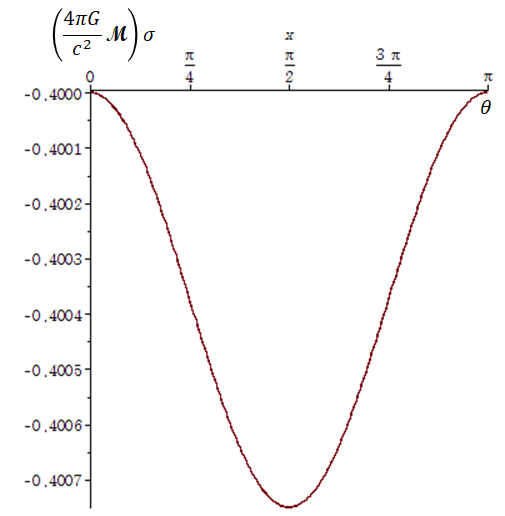}%
    }
  \caption{density distribution of thin-shell model in $\theta$ at constant $\beta=2, \alpha=0.2$}
  \end{figure}


\section{perturbation of symmetric rotating body}

Let a rigid body with principal moments $I_1, I_2,$ and $I_3$ rotate around $x_{\mathrm{3}}$-axis ($\boldsymbol{\omega }\mathrm{=}{\omega }_{\mathrm{3}}{\boldsymbol{e}}_{\mathrm{3}}$) and we apply a small perturbation, the angular velocity vector assumes the form as
\[\boldsymbol{\omega }\mathrm{=}\lambda {\boldsymbol{e}}_{\mathrm{1}}+\mu {\boldsymbol{e}}_{\boldsymbol{\mathrm{2}}}+{\omega }_{\mathrm{3}}{\boldsymbol{e}}_{\mathrm{3}},\]
where $\lambda $ and $\mu $ are small quantities.

The Euler equations becomes
\Be
&&(I_2-I_3)\mu \omega_3 - I_1 \dot{\lambda} =0, \nonumber \\
&&(I_3-I_1)\lambda \omega_3 - I_2 \dot{\mu} =0, \nonumber \\
&&(I_1-I_2)\lambda \mu - I_3 \dot{\omega}_3 =0.
\Ee
when $I_1 =I_2$, $\omega_3=$cosnt and
\Be
 \lambda(t) &=& A'e^{i\Omega t} + B'e^{-i\Omega t}, \\
\mu(t) &=& C'e^{i\Omega t} + D'e^{-i\Omega t},
\Ee
where
\[
\Omega=\omega_3\sqrt{\frac{(I_3-I_1)(I_3-I_2)}{I_1I_2}}=\omega_3\frac{I_3-I_1}{I_1}=
\omega_3\epsilon
\]
when $I_1=I_2<I_3$ and $\epsilon =\frac{I_3-I_1}{I_1}$ is the ellipticity of the body.
Here $A',B',C'$ and $D'$ are constants.
Thus we can say that the perturbation makes arise the two dimensional harmonic motion as
\Be
\omega_1(t)&=&\lambda(t)= A \cos(\Omega t-\alpha),\\
\omega_2(t)&=&\mu(t)=B \cos(\Omega t-\beta),
\Ee
with small constants $A$ and $B$.

\begin{figure}[t]
  \centering
    \subfigure
    {%
    \includegraphics[width=0.3\textwidth]{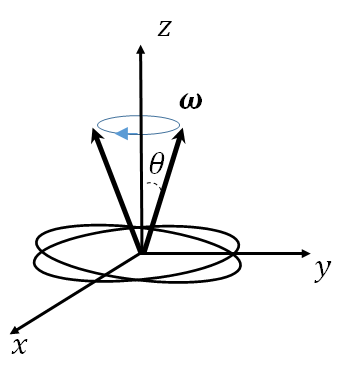}%
    }
  \caption{Circular and elliptic precessions around $z$-axis. The tip of $\boldsymbol{\omega}$ moves in a circle and an ellipse on the plane parallel to $x$-$y$ plane.   The former case has time-independent angle $\theta$ from $z$-axis and the latter has time-dependent $\theta$.}
\end{figure}

If $A=B$ and the phase difference is $\alpha-\beta=\pi/2$, it is a circular motion around $x_3$-axis in fixed frame, which is precession motion
and the angle $\theta=\theta_0$ is constant. See the Fig.~2. Thus we need the transformation
$\mathcal{R}_2\equiv\mathcal{R}_x(\theta_0)\mathcal{R}_z(\Omega t)$ from $\mathbb{I}_\mathrm{b}$
to $\mathbb{I}_\mathrm{f}$ and
we get the consequent components of $\mathbb{I}_\mathrm{f}$ as
\Be
I_{11} &=& I_1 (\cos^2\Omega t + \cos^2\theta_0 \sin^2\Omega t)+I_3 \sin^2\theta_0 \sin^2\Omega t, \nonumber\\
I_{12} &=&  (I_1-I_3)\sin^2\theta_0 \sin \Omega t \cos \Omega t, \nonumber \\
I_{13} &=& -(I_1-I_3)\sin\theta_0 \cos \theta_0 \sin \Omega t, \nonumber \\
I_{21} &=& (I_1-I_3)\sin^2\theta_0 \sin \Omega t \cos \Omega t, \nonumber \\
I_{22} &=&  I_1(\sin^2\Omega t+ \cos^2\theta_0 \cos^2\Omega t)+I_3 \sin^2\theta_0 \cos^2\Omega t, \nonumber\\
I_{23} &=& (I_1-I_3)\sin^2\theta_0 \sin \Omega t \cos \Omega t, \nonumber \\
I_{31} &=& -(I_1-I_3)\sin\theta_0 \cos \theta_0 \sin \Omega t, \nonumber \\
I_{32} &=&  (I_1-I_3)\sin\theta_0 \cos \theta_0 \cos \Omega t, \nonumber \\
I_{33} &=&  I_1 \sin^2\theta_0+I_3\cos^2 \theta_0. \nonumber 
\Ee

The second derivatives of the inertia tensor with respect to time are
\begin{widetext}
\be
\ddot{I}(t) =  \f{1}{2} \Omega^2 (I_3-I_1) 
\left(
  \begin{array}{ccc}
   4\sin^2\theta_0 \cos 2\Omega t & 4\sin^2\theta_0 \sin 2\Omega t  &  -\sin 2\theta_0 \sin \Omega t \\
    4\sin^2\theta_0 \sin 2\Omega t & -4\sin^2\theta_0 \cos 2\Omega t &  \sin 2\theta_0 \cos \Omega t \\
   -\sin 2\theta_0  \sin \Omega t & \sin 2\theta_0  \cos \Omega t & 0 
  \end{array}
\right) \label{M0} 
\ee
\end{widetext}

\begin{figure}[b]
  \centering
    \subfigure
    {%
    \includegraphics[width=0.4\textwidth]{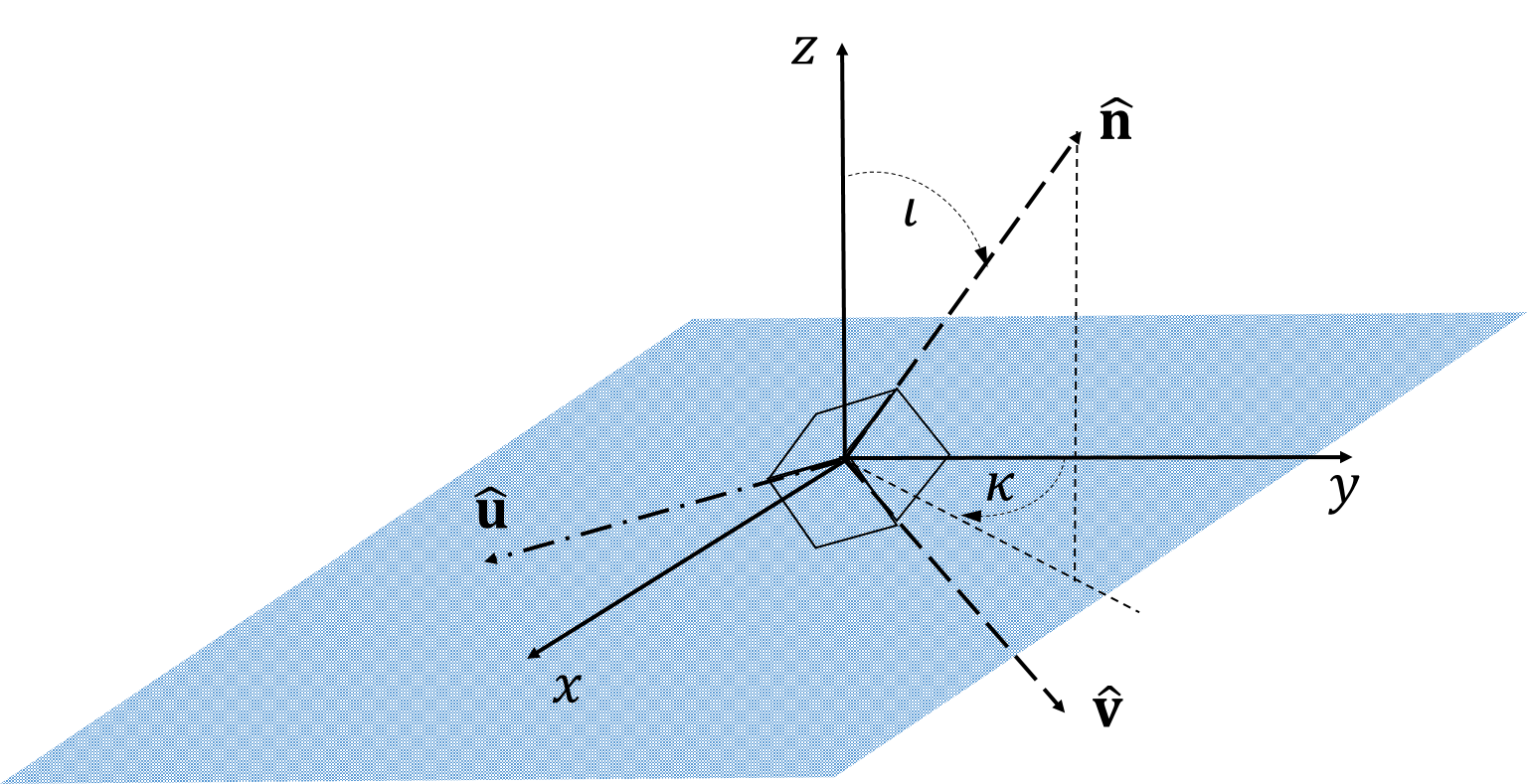}%
    }
  \caption{The relation between $(\hbrx,\hbry,\hbrz)$ frame and the $(\hbru,\hbrv,\hbrn)$ frame. The vector $\hbrn$ is at an angle $\iota$ from $z$-axis and $\kappa$ from $y$-axis \cite{MM}.}
  \end{figure}

To calculate the amplitudes for a gravitational waves propagating in the generic direction $\hbrn$, we introduce two unit vectors $\hbru$ and $\hbrv$ orthogonal to $\hbrn$ so that $\hbru \times \hbrv = \hbrn $ as shown in Fig.~3.

Therefore the strains $h_+$ and $h_\times$ in direction $\hbrn$ with inclination angle $\iota$ are
\Be
h_+(t;\iota) &=& -\f{G}{rc^4}( \ddot{I}_{11}-\ddot{I}_{22}\cos^2\iota - \ddot{I}_{33}\sin^2 \iota +
\ddot{I}_{23}\sin 2\iota \nonumber \\
&\simeq& h_0 [2 \theta_0^2 (1+\cos^2\iota)\cos 2\Omega t + 2\theta_0 \sin\iota \cos\iota \cos \Omega t], \nonumber \\
\\
h_\times(t;\iota)
&=& -\f{G}{rc^4}( 2\ddot{I}_{12}\cos\iota - 2\ddot{I}_{13}\sin\iota) \nonumber \\
&=& h_0 ( 4\theta_0^2 \cos\iota \sin 2\Omega t + 2\theta_0 \sin\iota \sin\Omega t ),
\Ee
where
\[
h_0 = -\f{G}{rc^4}I_1\epsilon\Omega^2.
\]
The results are the same as the free precession case of symmetric body \cite{ZS}  except that $\theta_0 = \sqrt{2}A/\omega_3$ and $\Omega=\epsilon\omega_3$.
That is, the amplitude depends on the perturbation size and the frequency depends on the
angular velocity of the rotation.
When $\iota=0$,
\[
h_+(t;0) = 4h_0 \theta_0^2 \cos 2\Omega t, \qquad  h_\times(t;0) = 4h_0 \theta_0^2 \sin 2\Omega t.
\]
They are same spectrum of free precession cases \cite{ZS} and the order of magnitude is $\epsilon^3\theta_0^2$.
When $\iota=\pi/2$,
\[
h_+(t;\pi/2) = 2h_0 \theta_0^2 \cos 2\Omega t, \qquad  h_\times(t;\pi/2) = 2h_0 \theta_0 \sin \Omega t.
\]
In this case, the size of $h_+(t;\pi/2)$ is smaller than the size of $h_\times$ in $\Delta$, and is half of $h_+(t;0)$. However, only the $\Omega$ spectrum appears in $h_\times(t;\pi/2)$.
The power radiated is
\Be
P &=& \f{2G}{5c^5}I_1^2\epsilon^2\Omega^6 \sin^2\theta_0 (\cos^2\theta_0+16\sin^2\theta_0) \nonumber \\
 &\simeq & \f{2G}{5c^5}I_1^2\epsilon^8\omega_3^6 \theta_0^2.  \label{RadPrec}
\Ee
for small angle $\theta_0$. It depends on $\epsilon^8$ and $\theta_0^2$.
The change rate of the angular momentum change $L$ is
\Be
\f{dL}{dt} &=& - \f{2G}{5c^5}I_1^2\epsilon^2\Omega^5 \sin^2\theta_0 (\cos^2\theta_0+16\sin^2\theta_0) \nonumber \\
 &\simeq & - \f{2G}{5c^5}I_1^2\epsilon^7\omega_3^5 \theta_0^2. \label{J_Prec}
\Ee
for small angle $\theta_0$. Therefore
\[
\f{dJ}{dt}=\f{dE}{dt}\f{1}{\Omega} = \f{dE}{dt}\f{1}{\epsilon\omega_3}.
\]
Where $J$ is the loss of angular momentum.

\section{gravitational waves by thin-shell wormhole}

We need the inertia tensor for the rotating wormhole, because of the consideration of the rigid body.
Note that the two-dimensional surface $t=$ const, $r=b$ in Kerr spacetime is actually an ellipsoid of revolution having minor and major axis equal to $b$ and $\sqrt{b^2+a^2}$.
In order to calculate the moment of inertia for thin shell, we need the surface integral over the ellipsoid. In the surface integral, the oblate factor is $ b\sqrt{b^2+a^2}\sqrt{1+\zeta \cos^2\theta} \sin\theta d\theta = {\cal{M}} \rho_\beta \sqrt{b^2+a^2}\sin\theta d\theta $ instead of $b^2\sin\theta d\theta $ for sphere. Here $\zeta = a^2/b^2=\alpha^2/\beta^2$.

To see the simple estimation of mass dependence on $J$, there is a relation like as
\cite{HKL2017}
\be
{\cal J}= {\cal M}^2 a_*, \label{AM_Mass}
\ee
where $a_*$ is the dimensionless spin parameter with value between 0 and 1. We can set $\beta=4 >2$.
The assumption of smallness of $\f{\alpha}{\beta} \simeq \f{a_*}{4}$ is validated for any mass.

Therefore, the moment of inertia for ellipsoid is
\be
I_{ij}=\int^{2\pi}_0\int^\pi_0[r^2\delta_{ij}-x_ix_j]\sigma {\cal{M}}\rho_\beta \sqrt{b^2+a^2} \sin \theta d \theta d \phi.
\ee
The $\sigma$ is used instead of $\sigma_0$ in calculating inertia tensor, because of the using  $T^{\hat{0}\hat{0}}$  for gravitational waves.
Here the $\cal{M}\rho_\beta \sigma $ is expanded for small $\zeta $ as
\[
{\cal{M}}\rho_\beta \sigma \cong -\frac{1}{2\pi}\frac{\sqrt{(\beta^2-2\beta)}}{\beta }[ 1 + \zeta ( P + Q x^2)],
\]
where
\[
x=\cos\theta, \qquad P = \frac{\beta +2}{2\beta(\beta -2)}, \qquad
Q = -\frac{1}{2}+\frac{3}{2\beta}.
\]

With this formula we get the components of moment
\Be
I_1 &=& I_2 = I_{xx} = \sigma_0 S_0 \frac{2}{3} b^{2} \left[ 1 + \zeta \left( P + 1 + \frac{2}{5}Q \right)\right] \nonumber \\
&=& - \f{4}{3}b^3 \sqrt{1-\f{2}{\beta}} \left[ 1 + \zeta \left( \frac{16\beta^2-10\beta -4}{20\beta (\beta -2 )}\right)\right]\left(\frac{c^2}{G}\right),
\\
I_3 &=& I_{zz} = \sigma_0 S_0 \frac{2}{3} b^{2} \left[ 1 + \zeta \left(P+\frac{3}{2}+\frac{1}{10}Q\right)\right] \nonumber \\
&=& - \f{4}{3}b^3 \sqrt{1-\f{2}{\beta}}\left[ 1 + \zeta \left(\frac{29\beta^2-45\beta+14}{20\beta(\beta-2)}
\right)\right]\left(\frac{c^2}{G}\right),
\Ee
where $S_0$ is the surface area of sphere of radius $b$ when $a$ is zero. These go to sphere case in the limit of $\zeta \to 0$
\[
I_3-I_1={\sigma }_0\frac{\mathrm{8}}{\mathrm{3}}\pi b^{\mathrm{4}}\zeta \frac{\mathrm{13}{\beta }^{\mathrm{2}}-\mathrm{35}\beta +\mathrm{18}}{\mathrm{20}\beta \mathrm{(}\beta -\mathrm{2)}}\mathrm{=}{\sigma }_0\frac{\mathrm{8}}{\mathrm{3}}\pi b^{\mathrm{4}}\zeta \frac{\mathrm{(13}\beta -\mathrm{9)\ }}{\mathrm{20}\beta }\].
The ellipticity is
\be
\epsilon =\frac{I_{\mathrm{3}}-I_{\mathrm{1}}}{I_{\mathrm{1}}}\mathrm{=}\frac{\zeta \frac{\mathrm{(13}\beta -\mathrm{9)\ }}{\mathrm{20}\beta }}{\mathrm{1}+\zeta \mathrm{(}\frac{\mathrm{16}{\beta }^{\mathrm{2}}-\mathrm{10}\beta -\mathrm{4}}{\mathrm{20}\beta \mathrm{(}\beta -\mathrm{2)}}\mathrm{)}}\cong \zeta \frac{\mathrm{(13}\beta -\mathrm{9)\ }}{\mathrm{20}\beta }.
\ee

The amplitude of the strain due to the dominant term in the value of $I_1$ is
\Be
h&=&|h_0|\theta_0^2=\frac{4G}{c^4r}\epsilon |I_1|\Omega^2\theta_0^2 \nonumber \\
&=& \frac{16}{{3c^2r}}b^3 \sqrt{1-\f{2}{\beta}} \left( \f{13\beta-9}{20\beta}\right)^3 \zeta^3 \omega^2 \theta_0^2.
\Ee
It is proportional to $b^3$ and $\zeta^3$, the consequent dependencies are $a^6b^{-3}$.
If we have reasonable $\beta =4$
\be
h\sim {10}^{-32}{\left(\frac{r}{{10}^8\mathrm{ly}}\right)}^{-1}{\left(\frac{b}{600\mathrm{km}}\right)}^3
{\left(\frac{\zeta}{10^{-2}}\right)}^3{\left(\frac{\theta_0}{10^{-2}}\right)}^2
{\left(\frac{\omega}{100\mathrm{Hz}}\right)}^2\nonumber \\
\ee
for $M=100M_\odot$.
This amplitude is very weak to be detected by present
gravitational wave observers. Since $\sigma$ is negative unlike normal objects, the wave has opposite sign, meaning a phase shift with $\pi$.

 The luminosity is
\Be
L_{GW} &=& \frac{2}{5}\frac{G}{c^5}I_1^2\epsilon^8\theta_0^2\omega^6  \nonumber \\
&=& \frac{32}{45Gc}b^6 \left( 1-\f{2}{\beta} \right) \left( \f{13\beta-9}{20\beta}\right)^8 \zeta^8 \omega^6 \theta_0^2,
\Ee
in terms of the dominant term.
This formula shows that the luminosity is proportional to the $a^{16}b^{-10}$.
\[
L_{GW}\sim {10}^{25}{\left(\frac{b}{600\mathrm{km}}\right)}^6{\left(\frac{\zeta}{10^{-2}}\right)}^8
{\left(\frac{\theta_0}{10^{-2}}\right)}^2
{\left(\frac{\omega}{100\mathrm{Hz}}\right)}^6
 \]
for $M=100M_\odot$ and $\beta=4$.  The energy radiation rate looks very large so that it radiates away in very short time. It seems that the precessing wormhole is very unstable.
However, the effective mass is calculated by
\Be
m &=& \int^{2\pi}_0\int^\pi_0\sigma {\cal{M}}\rho_\beta \sqrt{b^2+a^2} \sin \theta d \theta d \phi   \nonumber \\
&=&  \sigma_0 S_0 \left[ 1 + \zeta^2 \left(
\f{\beta-1}{3(\beta-2)} \right) \right]
\nonumber \\
&=& -2b \sqrt{1-\f{2}{\beta}} \left[ 1 + \zeta^2 \left(
\f{\beta-1}{3(\beta-2)} \right) \right] \left(\f{c^2}{G}\right).
\Ee
The size of the mass is roughly $10^{50}$.
Therefore the time to radiate all mass energy is
\[
\f{mc^2}{L_{GW}} \simeq 10^{25}
{\left(\frac{b}{600\mathrm{km}}\right)}^{-5}{\left(\frac{\zeta}{10^{-2}}\right)}^{-8}
{\left(\frac{\theta_0}{10^{-2}}\right)}^{-2}
{\left(\frac{\omega}{100\mathrm{Hz}}\right)}^{-6},
 \]
which is much longer than the cosmic time to radiate whole matter out.

The angular momentum loss rate is
\Be
\f{dL}{dt} &\simeq & - \f{2G}{5c^5}I_1^2\epsilon^7\omega_3^5 \theta_0^2 \nonumber \\
&=&  - \frac{32}{45Gc} b^6 \left( 1-\f{2}{\beta} \right) \left( \f{13\beta-9}{20\beta}\right)^7 \zeta^7 \omega^5 \theta_0^2.
\Ee

\section{Summary and Discussion}

We consider the toy model of gravitational waves generation by and thin-shell wormhole to consider the relationship of gravitational waves with wormhole, by assuming the ‘slowly’ rotating case under feasible conditions, such as mass limit.

We examined the nature of energy density to see the differences from gravitational waves by normal matter. We can see the signatures (negative property) of the density and they do not affect on the components of quadrupole moment.

We found the amplitude of the wave, gravitational waves luminosity, life time of wormhole in terms of throat size and angular momentum. The order of magnitude is very small, comparing to the binary merger cases. The gravitational wave spectrum
is determined by the rotation angular velocity of the thin-shell wormhole.
The smallness of all gravitational waves related quantities comes from ellipticity or the fractional ratio of angular momentum per mass to the wormhole throat size.

\acknowledgments
This work was supported by National Research Foundation of Korea (NRF) funded by the Ministry of
Education (2021R1I1A1A01056433).


\begin{thebibliography}{99}

\bibitem{Aasi} J. Aasi et al. Class. Quant. Grav. {\bf 32}, 074001 (2015).

\bibitem{Ace} F. Acernese et al. Class. Quant. Grav. {\bf 32}, 024001 (2015).

\bibitem{TBBL2021} A. Toubiana, S. Babak, E. Barausse, and L. Lehner, Phys. Rev. D {\bf 103}, 064042 (2021).

\bibitem{ATKAA} F. Abe, Asrtophys. J. {\bf 725}, 787 (2010); Y. Toki, T. Kitamura, H. Asada, and F. Abe, Asrtophys. J. {\bf 740}, 121 (2011).

\bibitem{NTY2013} P. G. Nedkova, V. N. Tinchev, and S. S. Yazadjiev, Phys. Rev. D {\bf 88}, 124019 (2013).

\bibitem{THY2012} N. Tsukamoto, T. Harada, and K. Yajima, Phys. Rev. D {\bf 86}, 104062 (2012).

\bibitem{SWK92} Sung-Won Kim, Phys. Rev. D {\bf 46}, 2428 (1992).

\bibitem{Ellis} H. G. Ellis, J. Math. Phys. {\bf 14}, 104 (1973).

\bibitem{Bron} K. A. Bronnikov, Acta Phys. Polon. B 4, 251 (1973).

\bibitem{MT}
M.S. Morris and K.S. Thorne, Am. J. Phys. {\bf 56}, 395 (1988).

\bibitem{Visser} M. Visser, Nucl. Phys. B{\bf 328}. 203 (1989).

\bibitem{HV} D. Hochberg and M. Visser, Phys. Rev. D {\bf 56}, 4745 (1997).

\bibitem{Ov} A. Ovgun, Eur. Phys. J. Plus {\bf 131}. 389 (2016).

\bibitem{Teo} E. Teo, Phys. Rev. D {\bf 58}, 024014 (1998).


\bibitem{KS2008} P. E. Kashargin and S. V. Sushkov, Gravit. Cosmol. {\bf 14}, 80 (2008); Phys. Rev. D {\bf 78}, 06071 (2008).




\bibitem{KK2014} B. Keihaus and J. Kunz, Phys. Rev. D {\bf 90}, 121503 (2014).

\bibitem{CKK2016} X. Y. Chew, B. Keihaus, and J. Kunz, Phys. Rev. D {\bf 94}, 104031 (2016).

\bibitem{Visser_book} M. Visser, Lorentzian Wormholes (Springer-Verlag, New York, 1996), pp. 75-78.

\bibitem{KS11}
P. E. Kashargin and S. V. Sushkov, Gravit. Cosmol. {\bf 17}, 119 (2011).












\bibitem{Mechanics}  S. T. Thornton and J. B. Marion, {\em Classical Dynamics of Particles and Systems}, 5ed., Thomson/Brooks/Cole, Belmont, CA (2003).


\bibitem{MM} M. Maggiore, {\em Gravitational Waves, Volume 1: Theory and Experiments}, Oxford University Press: New York (2008).

\bibitem{ZS} M. Zimmermann and E. Szedenits, Jr., Phys. Rev. D {\bf 20} 351 (1979).


\bibitem{HKL2017} T. Harko, Z. Kov\'{a}cs, and F.S.N. Lobo, Phys. Rev. D{\bf 79}, 064001 (2009).



\end{thebibliography}
\end{document}